\documentclass[3p,times,procedia,secnumarabic, graphics,floatfix, nofootinbib,tightenlines,nobibnotes, aps, prl, 12pt]{elsarticle}

\usepackage{graphicx}
\usepackage{amsmath,amssymb}
\usepackage{amsfonts}
\usepackage{multirow}

\biboptions{authoryear}

\begin{document}

\begin{frontmatter}

\title{Effect of stochastic transition in the fundamental diagram of traffic flow}
\author[a]{Adriano F. Siqueira}
\author[a]{Carlos J. T. Peixoto}
\author[b]{Chen Wu}
\author[a,c]{Wei-Liang Qian\corref{cor1}}

\address[a]{Departamento de Ci\^encias B\'asicas e Ambientais, Escola de Engenharia de Lorena, Universidade de S\~ao Paulo, SP, Brasil}
\address[b]{Shanghai Institute of Applied Physics, Shanghai, China}
\address[c]{Departamento de F\'isica e Qu\'imica, Faculdade de Engenharia de Guaratinguet\'a, Universidade Estadual Paulista, SP, Brasil}

\date{Jan, 03, 2016}

\begin{abstract}
In this work, we propose an alternative stochastic model for the fundamental diagram of traffic flow with minimal number of parameters.
Our approach is based on a mesoscopic viewpoint of the traffic system in terms of the dynamics of vehicle speed transitions.
A key feature of the present approach lies in its stochastic nature which makes it possible to study not only the flow-concentration relation, namely, the fundamental diagram, but also its uncertainty, namely, the variance of the fundamental diagram \textemdash an important characteristic in the observed traffic flow data.
It is shown that in the simplified versions of the model consisting of only a few speed states, analytic solutions for both quantities can be obtained, which facilitate the discussion of the corresponding physical content.
We also show that the effect of vehicle size can be included into the model by introducing the maximal congestion density $k_{max}$.
By making use of this parameter, the free flow region and congested flow region are naturally divided, and the transition is characterized by the capacity drop at the maximum of the flow-concentration relation.
The model parameters are then adjusted to the observed traffic flow on the I-80 Freeway Dataset in the San Francisco area from the NGSIM program, where both the fundamental diagram and its variance are reasonably reproduced.
Despite its simplicity, we argue that the current model provides an alternative description for the fundamental diagram and its uncertainty in the study of traffic flow.
\end{abstract}

\cortext[cor1]{wlqian@usp.br}


\end{frontmatter}


\section{Introduction}
Aside from its complexity and nonlinearity, traffic flow modeling has long attracted the attention of physicists due to the connections to transport theory and hydrodynamics (For reviews, see for example \citep{traffic-flow-review-07,traffic-flow-review-01,traffic-flow-btz-02,traffic-flow-review-02,traffic-flow-review-03,traffic-flow-review-04}).
Corresponding to the three main scales of observation in physics, traffic flow models can generally be categorized into three classes,  
namely, microscopic, mesoscopic and macroscopic approaches.
The macroscopic models \citep{traffic-flow-hydrodynamics-01,traffic-flow-hydrodynamics-02,traffic-flow-hydrodynamics-03,traffic-flow-hydrodynamics-04,traffic-flow-hydrodynamics-05,traffic-flow-hydrodynamics-06,traffic-flow-hydrodynamics-07,traffic-flow-hydrodynamics-08,traffic-flow-hydrodynamics-09,traffic-flow-hydrodynamics-16,traffic-flow-hydrodynamics-17} describe the traffic flow at a high level of aggregation, where the system is treated as a continuous fluid without distinguishing its individual constituent parts.
In this approach, the traffic stream is represented in terms of macroscopic quantities such as flow rate, density and speed.
Many methods in the conventional hydrodynamics thus can be directly borrowed into the investigation of traffic flow. 
For instance, one may discuss shock waves \citep{traffic-flow-hydrodynamics-01,traffic-flow-hydrodynamics-02}, the stability of the equation of motion \citep{traffic-flow-hydrodynamics-04,traffic-flow-hydrodynamics-07}, or investigate the role of viscosity \citep{traffic-flow-hydrodynamics-08} analogous to those for real fluid. 
Mathematically, the problem is thus expressed in terms of a system of partial differential equations.
The microscopic approach, on the other hand, deals with the space-time behavior of each individual vehicle as well as their interactions at the most detailed level.
In this case, an ordinary differential equation is usually written down for each vehicle. 
Due to its mathematical complexity, approximation is commonly introduced in order to obtain asymptotic solutions or to make the problem less computationally expensive.
The car-following models \citep{traffic-flow-micro-01,traffic-flow-micro-02,traffic-flow-micro-03,traffic-flow-micro-04,traffic-flow-micro-05,traffic-flow-micro-06,traffic-flow-micro-10,traffic-flow-micro-11}, optimal velocity models \citep{traffic-flow-micro-07,traffic-flow-micro-09,traffic-flow-micro-17,traffic-flow-micro-19,traffic-flow-micro-20}
and the cellular automaton \citep{traffic-flow-cellular-01,traffic-flow-cellular-02,traffic-flow-cellular-03,traffic-flow-cellular-04,traffic-flow-cellular-05} all can be seen as microscopic approaches in this context.
For certain cases, such as Greenberg's logarithmic model \citep{traffic-flow-hydrodynamics-03,traffic-flow-micro-02}, 
the above two approaches were shown to be equivalent in reproducing the fundamental diagram of traffic flow.
A mesoscopic model \citep{traffic-flow-btz-01,traffic-flow-btz-02,traffic-flow-btz-03,traffic-flow-btz-04} seeks compromise between the microscopic and the macroscopic approaches.
The model does not attempt to distinguish nor trace individual vehicles, instead, it describes traffic flow in terms of vehicle distribution density as a continuous function of time, spatial coordinates and velocities. 
The dynamics of the distribution function, following methods of statistical mechanics \citep{statistical-mechanics-huang}, is usually determined by an integro-differential equation such as the Boltzmann equation.
Most mesoscopic models are derived in analogy to gas-kinetic theory.
As it is known that hydrodynamics can be obtained by using the Boltzmann equation \citep{btz-hydrodynamics-01,btz-hydrodynamics-02,btz-hydrodynamics-03}, the mesoscopic model for traffic flow has also been used to obtain the corresponding macroscopic equations \citep{traffic-flow-hydrodynamics-06,traffic-flow-hydrodynamics-08}.
These efforts thus provide a sound theoretical foundation for macroscopic models, besides heuristic arguments and lax analogies between traffic flow and ordinary fluids.

One important empirical measurement for a long homogeneous freeway system is the so called ``fundamental diagram" of traffic flow. It is plotted in terms of vehicle flow $q$ as a function of vehicle density $k$:
\begin{eqnarray}
q=q(k)
\end{eqnarray}
In a macroscopic theory, when the dynamics of the system is determined by an Euler-like or Navier-Stokes-like equation of motion, the fundamental diagram can be derived.
Alternatively, one may use the fundamental diagram as an input together with the conservation of vehicle flow and the initial conditions to determine the temporal evolution of the system.
Also, the equation of motion of either the microscopic or the mesoscopic model can be employed to calculate the fundamental diagram.
The resulting theoretical estimations from any of the above approaches can then be used to compare to the empirical observations 
which have been accumulated on highways in different countries for nearly 8 decades (see for instance ref.\citep{traffic-flow-review-02,traffic-flow-data-01,traffic-flow-phenomenology-01}).
The following common features are observed in most of the data:
(1) Usually the flow-concentration curve is divided into two different regions of lower and higher vehicle density, which correspond to ``free" and ``congested" flow;
(2) The maximum of the flow occurs at the junction between the free and congested region and
(3) Congested flow in general presents a broader scattering of the data points on the flow-concentration plane, in comparison to that of the free flow. 
In other words, the variance of flow for free traffic flow is relatively small, it increases as the vehicle density increases,
and eventually the system becomes unstable or chaotic toward the onset of traffic congestion.
For this very reason, it is understood by many authors that the transition from free traffic to congestion is a {\it phase transition}. 
Most traffic flow models are able to reproduce the main features of the observed fundamental diagram;
in particular, traffic congestion is understood to be closely connected to the instability of the equation of motion \citep{traffic-flow-hydrodynamics-04,traffic-flow-micro-07,traffic-flow-micro-08,traffic-flow-micro-09,traffic-flow-micro-17}, 
or to the phase transition of the system \citep{traffic-flow-catastrophe-01,traffic-flow-catastrophe-02,traffic-flow-review-01,traffic-flow-cellular-02}.
On the other hand, uncertainty is also observed in the data which can be mostly expressed in terms of the variance of the fundamental diagram.
The latter has been an intriguing topic in the recent years \citep{traffic-flow-hydrodynamics-10,traffic-flow-data-03,traffic-flow-micro-12,traffic-flow-hydrodynamics-11,traffic-flow-review-05,traffic-flow-btz-05}.
In fact, methodologies involving stochastic modeling have aroused much attention,
either from the macroscopic viewpoint \citep{traffic-flow-phenomenology-02,traffic-flow-phenomenology-03,traffic-flow-phenomenology-04,traffic-flow-phenomenology-05},
from the microscopic aspect such as car-following and cellular automaton models \citep{traffic-flow-micro-13,traffic-flow-micro-14,traffic-flow-micro-15,traffic-flow-cellular-01,traffic-flow-cellular-05,traffic-flow-micro-16},
or from other phenomenological approaches \citep{traffic-flow-phenomenology-06,traffic-flow-phenomenology-07,traffic-flow-phenomenology-08} such as those introduce uncertainty directly into the fundamental diagram or road capacity.

The present work follows the above line of thought to explore the stochastic nature of traffic flow. 
First, we will employ a proper mathematical tool to tackle the problem.
One notes that a model simple in its mathematical form may not imply the most appropriate interpretation for an elementary physical system.
As it is well known, the random motion of particles suspended in a fluid, known as the Brownian motion, 
is best described by the Wiener process.
The latter involves the rules of stochastic calculus since the corresponding equation of motion, a stochastic differential equation (SDE), is typically not differentiable. 
Secondly, in our approach, one demands the model to be of microscopic/mesoscopic origin, 
meanwhile it shall not be subjected to special rules tailored for a specific traffic scenario or some certain experimental data,
so that the model could succeed in describing traffic evolution under many different traffic conditions.
Furthermore, one also requires the model to be simple enough so that analytic solutions may be obtained.
This motivated us to carry out the present study.
In this work, we introduce a simple mesoscopic model for the traffic flow by the method of SDE. 
The equation of motion of the model governs the temporal evolution of the distributions of vehicles among different speed states.
In addition to the conventional transition terms, stochastic transition is introduced in order to describe the stochastic nature of traffic flow.
We show in our model that analytic solutions can be obtained not only for the expected value of speed and traffic flow, but also for their variances. 
These analytic solutions are then compared to the empirical data.
In the next section, we introduce our transport model which features a speed spectrum and the corresponding transitions dynamics among different speed states. 
To show the essence of the approach, the model is then simplified to consider only two speed states.
The resulting two speed model is discussed in detail in section III, where we
derive the analytic solutions for the flow-concentration curve and its variance. 
One also takes into consideration the effect of finite vehicle size, which is done by explicitly introducing the maximal congestion density $k_{max}$ into the transition rate in the form of a suppression factor, so that all the vehicles are forced into the low speed state when vehicle density approaches this limit.
We also present analytic results of the model in the presence of this additional parameter.
The physical content of these solutions is discussed.
In order to compare to the data, a chi-square fitting is carried out for model parameters in section IV and 
the results are presented together with the public I-80 freeway dataset from the NGSIM program. 
The last section is devoted to the conclusion remarks and perspectives.

\section{A stochastic transport model with discrete speed spectrum}

Let us consider a section of highway where the spatial distribution of the vehicles is homogeneous. For simplicity, one only considers discrete values for speed, namely $v_1,v_2,\cdots,v_D$ and the number of vehicles traveling at speed state $v_i$ is denotes by $n_i$. In time, a vehicle with speed $v_i$ may transit to another state $v_j$ according to the following set of SDE \citep{stochastic-difeq-oksendal}
\begin{eqnarray}
\frac{dn_i}{dt} = \sum_{j=1}^D c_{ij} n_j + \sum_{j=1}^D s_{ij} \sqrt{n_j}w_j
\label{bten}
\end{eqnarray}
where the speed transition on the r.h.s. of the equation is a summation of two contributions: the deterministic and stochastic transitions measured by the transition rates $c_{ij}$ and $s_{ij}$, 
and one introduces some randomness by the white noise $w_j$, which is a random signal characterized by a featureless (namely, constant) power spectral density. 
When $j\ne i$, the coefficients $c_{ij}$ and $s_{ij}$ measure the rate a vehicle with speed $v_j$ will transit to another state with speed $v_i$.
In particular, the transition rate $c_{ij}$ with $v_j>v_i$ corresponds to the breaking of the vehicle when the driver with speed $v_j$ encounters a vehicle with a lower speed $v_i$ in front of him, thus the driver slows down the vehicle and transits to the speed state $v_i$. This transition rate is of the same content as the loss part of the collision term, $\Gamma^{-}_{ij}$, in the Boltzmann-like equation introduced in ref.\citep{traffic-flow-btz-01} by Prigogine and Andrews. Identical to their work, our approach also has the transition proportional to the occupation number of the initial state, $n_j$. Similarly, the transition rate $c_{ji}$ with $v_j>v_i$ corresponding to the case when the driver increases his speed when approached by some other vehicle with a higher speed $v_j$, and therefore it corresponds to the gain part of the collision term in a Boltzmann equation. 
When $j = i$, the coefficients $c_{ii}$ and $s_{ii}$ reflect the transition from state $i$ due to some internal causes (such as human error).
As a result, these transition coefficients are connected to the above-mentioned traffic system characteristics. It is also worth noting that in our model we did not introduce the relaxation term as in \citep{traffic-flow-btz-01}, 
because the analytic solution of the equation automatically converges to a stable solution as shown in Eq.(\ref{ebte2}).
It is also shown below in Eq.(\ref{normal2}) that these transition coefficients are not completely independent, neither are they necessarily to be constants. 
In the present approach, the differential formalism is to be understood in terms of the It\^o interpretation \citep{stochastic-difeq-oksendal}. 
The stochastic transition rate is taken to be proportional to $\sqrt{n_j}$, 
so that the stochastic transition weighs as much as the deterministic transition given the same occupation number \citep{stochastic-difeq-allen}.
It is obvious that for any stable solution, $n_i$ must be bounded from above and below. 

In our model, the measured traffic speed $v$ is defined by
\begin{eqnarray}
v & \equiv & \frac{\sum_i n_i v_i}{\sum n_i} \label{v-measure}
\end{eqnarray}
Consequently, the traffic flow $q$ is defined as the product of speed $v$ and vehicle density $k$ as follows
\begin{eqnarray}
q \equiv k v  \label{q-measure}
\end{eqnarray}

One notes that the transition coefficients $c_{ij}$ and $s_{ij}$ can be seen as the element in the $i$-th row and $j$-th column of $D \times D$ matrices $c$ and $s$ respectively. 
When one is only interested in the time evolution of the expected value of $n_i$, 
the stochastic transition terms can usually be ignored \footnote{see for example, 
Theorem 3.2.1 of \citep{stochastic-difeq-oksendal} for the discussions of this condition. In our case when $s_{ij}$ is deterministic, the condition reduces to that $s_{ij}$ 
must be square-integrable in time.} and consequently Eq.(\ref{bten}) can be written as
\begin{eqnarray}
\frac{dn}{dt}= c n 
\label{bten-matrix}
\end{eqnarray}
where $n=\begin{pmatrix}
n_1\\ 
n_2\\ 
\vdots \\ 
n_D
\end{pmatrix}$ is a column matrix and $c$ is the $D \times D$ transition matrix defined above. 
Now, for a closed road system (a system which satisfies periodic boundary condition), the total number of vehicles is conserved, i.e. 
\begin{eqnarray}
\sum_i {n_i} = const. = N 
\label{normal1}
\end{eqnarray}
By using Eq.(\ref{normal1}), it is straightforward to show that 
\begin{eqnarray}
\sum_i c_{ij} = 0 
\label{normal2}
\end{eqnarray}
for any state $j$, which means that the matrix $c_{ij}$ is singular. Therefore, one may explicitly express $n_D$ in terms of $n_i (i=1,\cdots,D-1)$ and rewrite the equations for $n_i (i=1,\cdots,D-1)$ in terms of
the first $D-1$ rows of $c_{ij}$, namely
\begin{eqnarray}
\frac{dn_i}{dt} = \sum_{j=1}^{D-1} \tilde{c}_{ij} n_j +c_{iD}N 
\label{btem}
\end{eqnarray}
where $\tilde{c}_{ij} \equiv c_{ij}-c_{iD}$. Similarly, one may again view $\tilde{c}_{ij}$ as the $i$-th row and $j$-th column of a $(D-1) \times (D-1)$ matrix $\tilde{c}$ and rewrite the equation as 
\begin{eqnarray}
\frac{d\tilde{n}}{dt}= \tilde{c} \tilde{n} +\tilde{n}_0 
\label{btem-matrix}
\end{eqnarray}
where $\tilde{n}=\begin{pmatrix}
n_1\\ 
n_2\\ 
\vdots \\ 
n_{D-1}
\end{pmatrix}$ and
$\tilde{n}_0=\begin{pmatrix}
c_{1D}N\\ 
c_{2D}N\\ 
\vdots \\ 
c_{(D-1)D}N
\end{pmatrix}$ 
are ($(D-1) \times 1$) column matrices. Usually, $\tilde{c}$ is non-singular, and the number of degrees of freedom of the system is thus $D-1$.

As an example, let us discuss the case where all the elements of the matrix $c$ are constant, which always can be seen as an approximation when a small perturbation is introduced around a steady state.
In this case, the problem is simply reduced to the diagonalization of the $(D-1)\times (D-1)$ matrix $\tilde{c}$. 
A necessary condition to have stable physical solutions is all the eigen-values of the matrix $\tilde{c}$ must be negative. 
This is because any positive eigen-value would imply that the number of vehicles of some state increases unboundedly in time, which is not physical due to the condition introduced in Eq.(\ref{normal1}). 
On the other hand, traffic congestion is known to be closely related to the instability of the equation of motion for a realistic traffic system.
This implies that the matrix $\tilde{c}$ cannot be constant for a realistic model, where it ought to possess some regions with positive eigen-values.
However, we will not pursue this matter any further in the present study.
Finally we note that it can be shown straightforwardly that this model is related to the discrete limit of the Boltzmann equation approach \citep{traffic-flow-btz-02}.

\section{A simplified model with two speed states}

In order to show that the present model may describe the main features of the observed traffic flow data, 
we proceed to discuss the most simple case where one considers only two speed states. 
The equation of motion of this simplified model reads
\begin{eqnarray}
\frac{dn_1}{dt} &=& -p_{11}n_1 + p_{12}n_2 N^\alpha  - \sqrt{p_{11}n_1}w_1+\sqrt{p_{12}n_2 N^\alpha}w_2 \nonumber \\
\frac{dn_2}{dt} &=& -p_{22}n_2 N^\alpha + p_{21}n_1  - \sqrt{p_{22}n_2 N^\alpha}w_2 + \sqrt{p_{21}n_1}w_1
\label{bte2}
\end{eqnarray}
where $w_1$ and $w_2$ are independent white noises. One has $n_1+n_2=N$, $p_{11}=p_{21}$ and $p_{22} = p_{12}$ due to the normalization condition, Eqs.(\ref{normal1}-\ref{normal2}). 
It is noted by some authors \citep{traffic-flow-phenomenology-10,traffic-flow-phenomenology-11} that the decelerating and accelerating is asymmetric: the breaking is usually more abrupt than the acceleration. 
In our model, the parameter $\alpha$ was introduced to take this fact into account, where $\alpha > 1$ will be taken.  
The form of Eq.(\ref{bte2}) is common in the application of SDE \citep{stochastic-difeq-oksendal,stochastic-difeq-allen,stochastic-difeq-app-01}.

One is allowed to obtain its analytic solutions due to the simplicity of the model.
First, by ignoring the stochastic transition terms, the solution of the expected value of $n_i$ reads
\begin{eqnarray}
{\textrm E}[n_1(t)] &=& \frac{p_{22}N^{\alpha+1}}{p_{11}+p_{22}N^\alpha}+\left(n_1(0)-\frac{p_{22}N^{\alpha+1}}{p_{11}+p_{22}N^\alpha}\right)e^{-(p_{22}N^\alpha+p_{11})t} \nonumber \\
{\textrm E}[n_2(t)] &=& N-{\textrm E}[n_1]=\frac{p_{11}N}{p_{11}+p_{22}N^\alpha}+\left(N-n_1(0)-\frac{p_{11}N}{p_{11}+p_{22}N^\alpha}\right)e^{-(p_{22}N^\alpha+p_{11})t} 
\label{ebte2}
\end{eqnarray}
where $n_1(0) \equiv n_1(t=0)$. The steady state solution is obtained by taking the limit $t\rightarrow \infty$
\begin{eqnarray}
\lim_{t\rightarrow \infty}{\textrm E}[n_1(t)] \equiv {\textrm E}[n_1^{(\infty)}] &=& \frac{p_{22}N^{\alpha+1}}{p_{11}+p_{22}N^\alpha} \nonumber \\
\lim_{t\rightarrow \infty}{\textrm E}[n_2(t)] \equiv {\textrm E}[n_2^{(\infty)}] &=& N-\lim_{t\rightarrow \infty}{\textrm E}[n_1]=\frac{p_{11}N}{p_{11}+p_{22}N^\alpha} \label{expectedsteady}
\end{eqnarray}

The measured vehicle speed $v$ in this case is
\begin{eqnarray}
v & = & \frac{n_1^{(\infty)} v_1+ n_2^{(\infty)} v_2}{n_1+n_2} 
\end{eqnarray}
and its expected value reads
\begin{eqnarray}
{\textrm {E}}[v] &=& \frac{\sum_i {\textrm E}[n_i^{(\infty)}] v_i}{\sum n_i} = \frac{{\textrm E}[n_1^{(\infty)}] v_1+{\textrm E}[n_2^{(\infty)}]v_2}{N} = \frac{p_{11}v_2+p_{22}v_1N^\alpha}{p_{11}+p_{22}N^\alpha}
\label{v-expectation}
\end{eqnarray}
In literature, the above result is usually expressed in terms of traffic flow $q$ as a function of the vehicle density $k$, with the latter being written as
\begin{eqnarray}
k = \frac{N}{L}
\end{eqnarray}
where $L$ is the length of the highway section in question. Therefore, by using Eq.(\ref{q-measure}) one obtains the expected value of traffic flow $q$
\begin{eqnarray}
{\textrm {E}}[q] = {\textrm {E}}[kv]=  \frac{p_{11}v_2 k+p_{22}v_1 L^\alpha k^{\alpha+1}}{p_{11}+p_{22}L^\alpha k^{\alpha}}
\label{fc-expectation}
\end{eqnarray}

One sees when $\alpha > 1$, the vehicles have a tendency to transition to the low speed state when the total number $N$ increases, which is consistent with common sense: the average speed tends to decrease when the load on the highway system becomes heavier. 

\begin{figure}[!htb]
\begin{tabular}{cc}
\begin{minipage}{200pt}
\centerline{\includegraphics*[width=8cm]{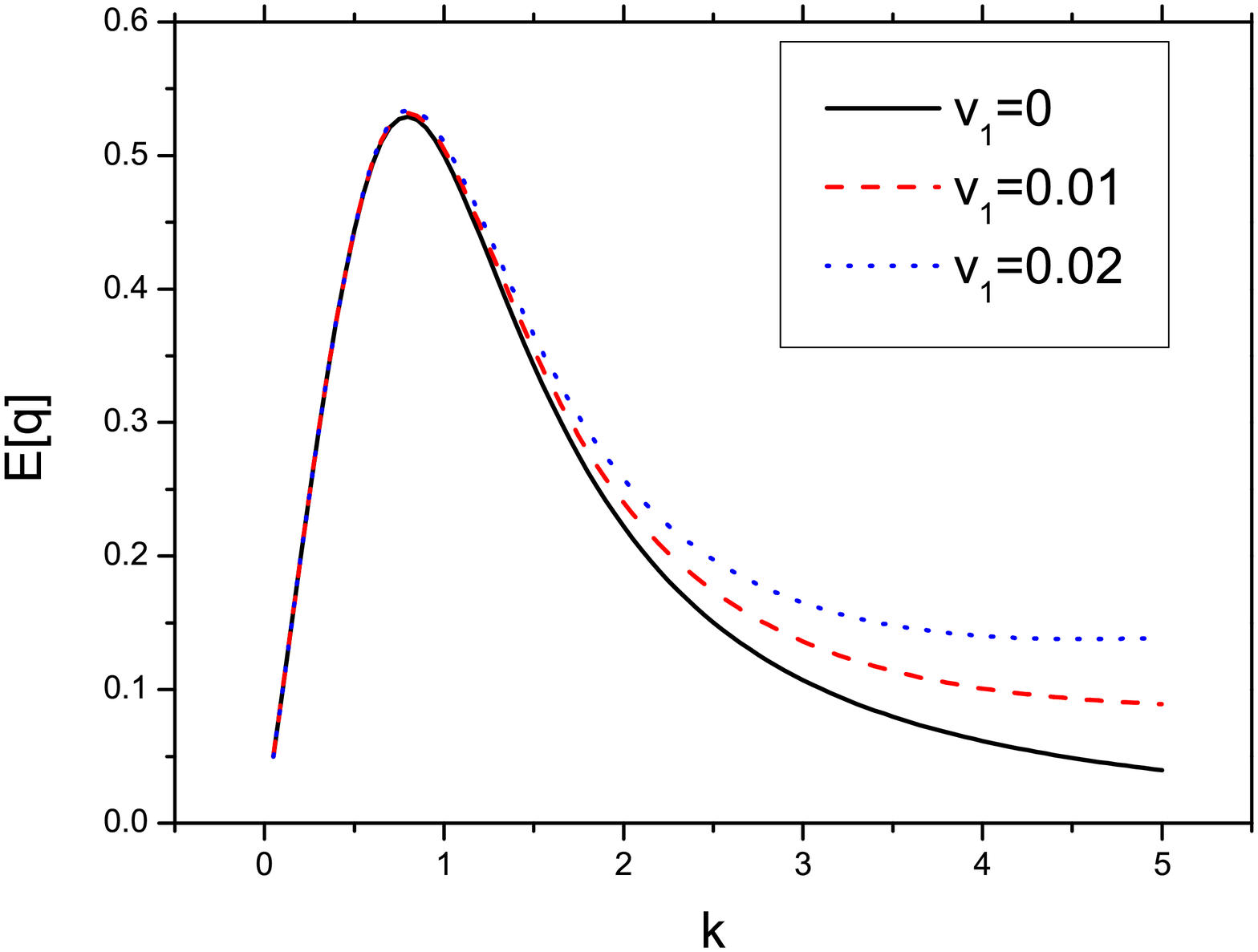} }
\end{minipage}
&
\begin{minipage}{200pt}
\centerline{\includegraphics*[width=8cm]{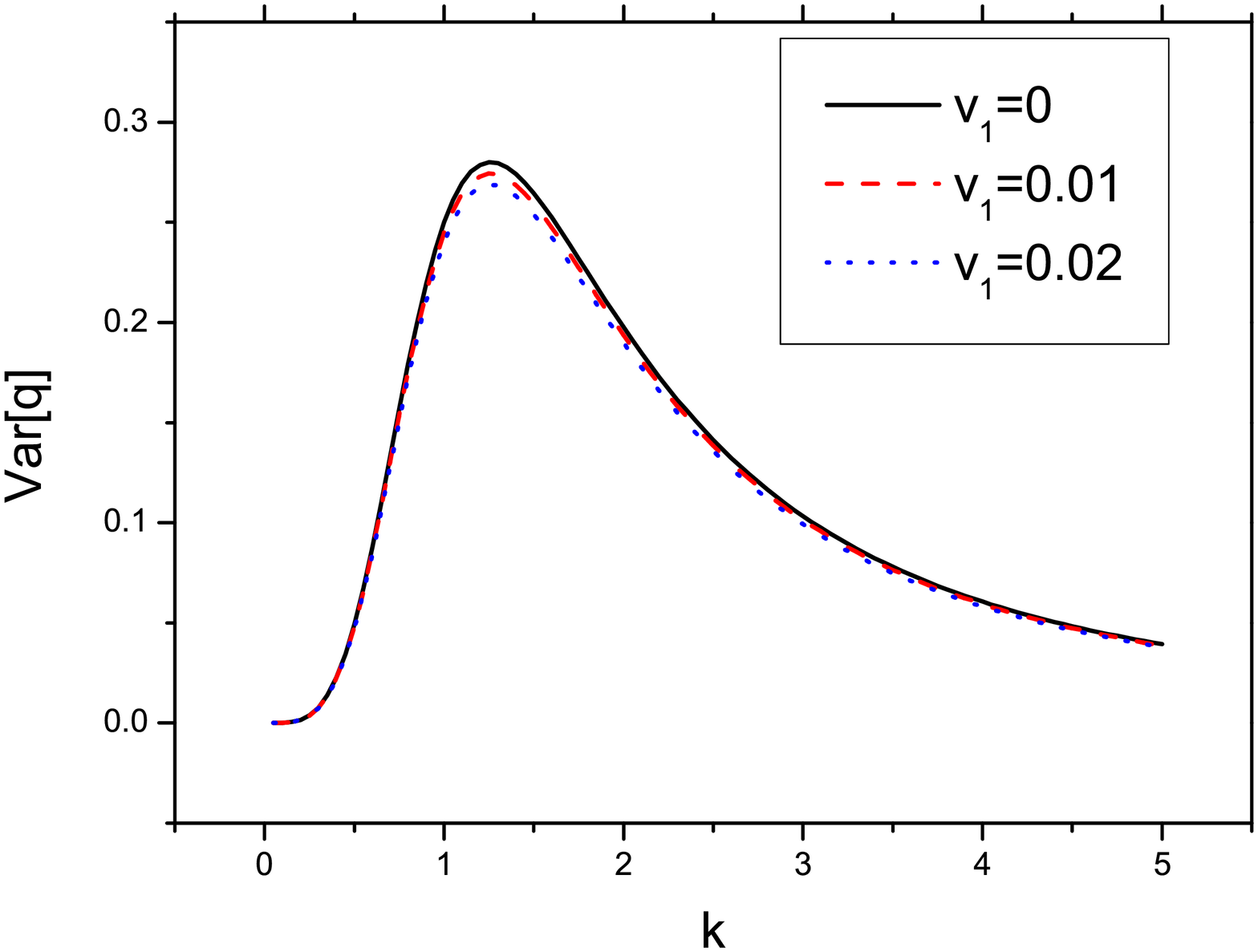} }
\end{minipage}
\\
\end{tabular}
\caption{Schematic fundamental diagram from the two-speed-state model where a trivial parameterization was adopted, namely,
$p_{11}=p_{22}=L=v_2=1$ and $v_1=0, 0.01, 0.02$. 
In the plot, $\alpha =3$ was used in order to reproduce the main shape of the fundamental diagram in most cases.
Left: the flow as a function of concentration, where the maximum of the flow appears at $k_c^{(1)} = \frac{1}{\sqrt[3]{2}} \sim 0.79$. Right: variance of the flow as a function of concentration, where the maximum appears at $k_c^{(2)} = {\sqrt[3]{2}} \sim 1.26$.}
\label{fc-model}
\end{figure}

In the left panel of Fig.\ref{fc-model}, we show a sketch plot of the resulting fundamental diagram of the two-speed-state model, obtained by assuming a set of rather trivial parameters in Eqs.(\ref{v-expectation}-\ref{fc-expectation}).
It is shown that the main feature of the flow concentration curve is naturally reproduced. 
The parameter $\alpha$ controls the shape of the curve, and it seems $\alpha \gtrsim 2$ gives qualitatively good agreement to the data which was also observed in many other cases \citep{traffic-flow-phenomenology-01,traffic-flow-data-01}.

As mentioned above, one important feature of the current approach is that the stochastic transitions introduce uncertainties in the number of vehicles, which thereafter cause the vehicle flow to fluctuate around its mean value.
The main features of the uncertainties of the traffic flow in data is well known \citep{traffic-flow-data-01,traffic-flow-phenomenology-01}.
In the present model we are able to calculate the variance of the flow-concentration curve analytically.
Following the standard procedure of It{\^o} calculus (see Appendix I for details), one finds the variance of the measured vehicle speed satisfies
\begin{eqnarray}
{\textrm {Var}}[q] = \frac{(v_1-v_2)^2}{L^2}\frac{p_{11}p_{22}L^{\alpha+1}k^{\alpha+1}}{(p_{11}+p_{22}L^\alpha k^\alpha)^2} \label{fc-variance}
\end{eqnarray}

As observed in the data, the variance of the speed is very small at small concentration.
This can be understood as follows, when there are very few vehicles on the highway, all of them tend to move at the upper speed limit, thus the variance is negligible.
In our model, for $\alpha \ge 0$, one finds that the variance goes to zero when $k \rightarrow 0$. 
If $\alpha < 0$, the variance will diverge at small concentration which makes the model unrealistic.
On the other hand, at very high density, all the vehicles tend to occupy the 
lowest speed state corresponding to $v_1$. 
It is easy to imagine that this corresponds to the case of a complete traffic jam, when all the vehicles are forced to stop, and consequently the variance of the speed also goes to zero. 
From Eq.(\ref{fc-variance}), one has the limit ${\textrm {Var}}[v] \rightarrow 0$, which is consistent with the above discussion. 

By employing Eq.(\ref{fc-expectation}) and taking $v_1=0$ and evaluating its derivative, one obtains the value of $k_c^{(1)}$ for maximal traffic flow
\begin{eqnarray}
k_c^{(1)} = \frac{1}{L} \left(\frac{1}{\alpha-1}\frac{p_{11}}{p_{22}}\right)^{1/\alpha}
\end{eqnarray}
Furthermore, the data indicates that the maximum of the flow variance corresponds to the onset of congestion, which usually occurs shortly after the maximum of the flow. 
From Eq.(\ref{fc-variance}), one obtains the corresponding value of $k_c^{(2)}$ for maximal variance
\begin{eqnarray}
k_c^{(2)} = \frac{1}{L} \left(\frac{\alpha+1}{\alpha-1}\frac{p_{11}}{p_{22}}\right)^{1/\alpha}
\end{eqnarray}
When $\alpha > 1$, one has $k_c^{(2)} > k_c^{(1)}$.
This is another motivation for our choice of the value of $\alpha$.
It is worth noting that the above features of our model come out quite naturally.
In the right panel of Fig.\ref{fc-model}, we show a sketch plot for the variance of the traffic flow in our model.

We also developed formulae for a simple version of the model with three speed states, the corresponding formulae can be found in Appendix II and numerical results are presented in the next section.

Up to this point, we have assumed that all vehicles have negligible sizes. As a matter of fact, the size of vehicles will not affect the free flow traffic state (FT), where almost all vehicles transit at the desired speed.
This is because in the free flow phase, the distance between successive vehicles is much bigger than the vehicle size. 
In the congestion flow region (HCT) of the fundamental diagram, the situation is quite different. As the vehicle density on the highway increases, the distance between vehicles decreases, and as a result, the effect of the vehicle size becomes more and more important. 
Not considering this effect, an immediate shortcoming of the above model is that according to Eq.(\ref{fc-expectation}), the traffic breaks down only at infinite vehicle density $k \rightarrow \infty$, which is obviously not realistic.
In reality, there exists a maximal congestion density $k_{max}$ whose value is more or less determined by the inverse of average vehicle size: since an automobile will not be able to physically packed into a density larger than it. The traffic flow shall completely break down at this limit density. To incorporate this effect into our model, we modify the breaking transition rate by the following substitution for the congested flow. 
\begin{eqnarray}
N^{\alpha} &\rightarrow& \beta N^{\alpha} \nonumber\\
\beta &=& N^{\alpha}\left(\frac{1}{1-\frac{k}{k_{max}}}\right)
\end{eqnarray}
The above subscription guarantees that the breaking transition rate goes to infinite when $k \rightarrow k_{max}$, in other words, the traffic breaks down completely at the density $k = k_{max}$.

It is not difficult to show that the corresponding results for the expected value and variance of the flow are modified as following:
\begin{eqnarray}
{\textrm {E}}[q^{(HCT)}] =   \frac{p_{11}v_2 k+p_{22}v_1 L^\alpha k^{\alpha+1}\beta}{p_{11}+p_{22}L^\alpha k^{\alpha}\beta} \\
\label{fc-expectation2}
{\textrm {Var}}[q^{(HCT)}] = \frac{(v_1-v_2)^2}{L^2}\frac{p_{11}p_{22}L^{\alpha+1}k^{\alpha+1}\beta}{(p_{11}+p_{22}L^\alpha k^{\alpha} \beta)^2} 
\label{fc-variance2}
\end{eqnarray}  

\begin{figure}[!htb]
\begin{tabular}{cc}
\begin{minipage}{200pt}
\centerline{\includegraphics*[width=8cm]{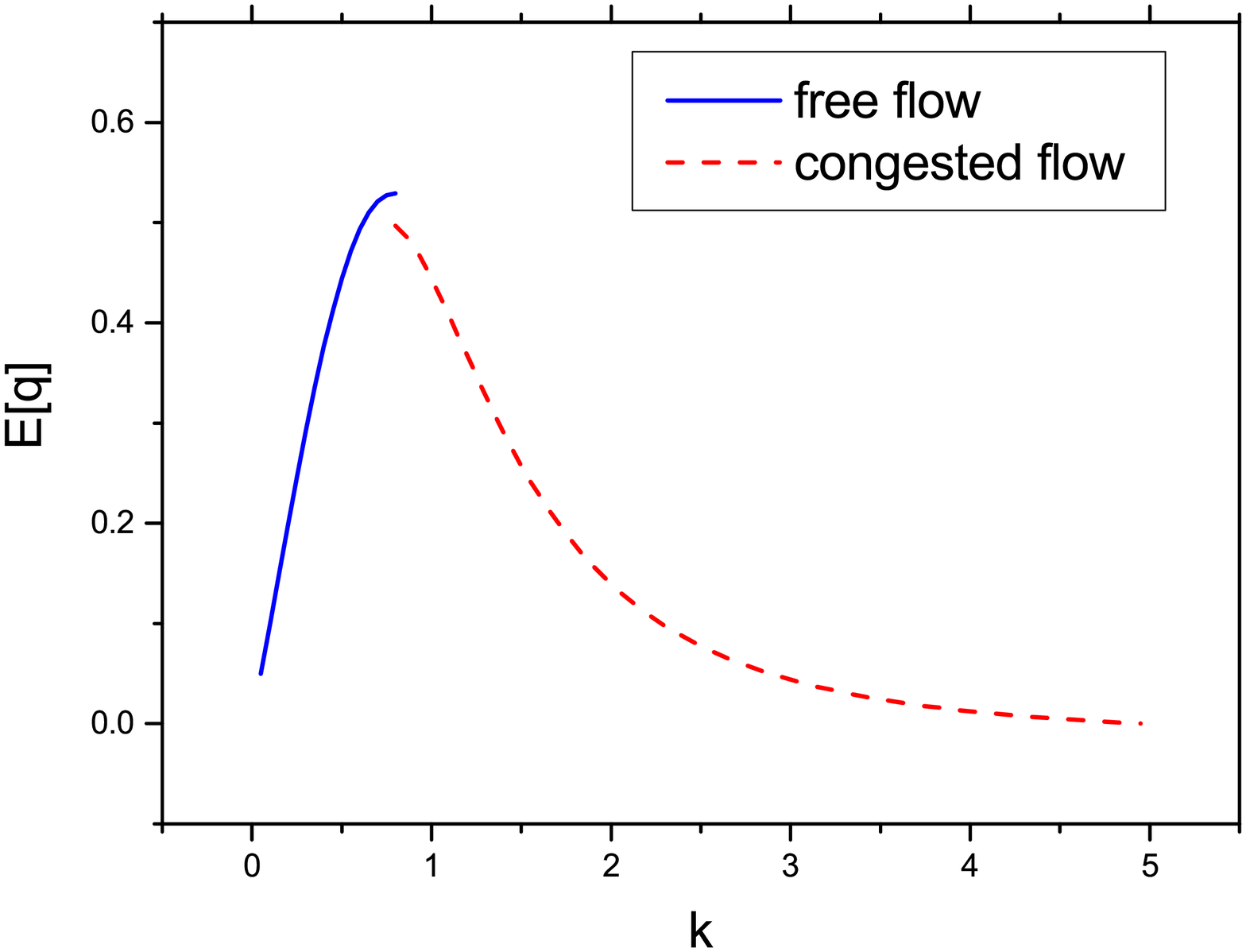} }
\end{minipage}
&
\begin{minipage}{200pt}
\centerline{\includegraphics*[width=8cm]{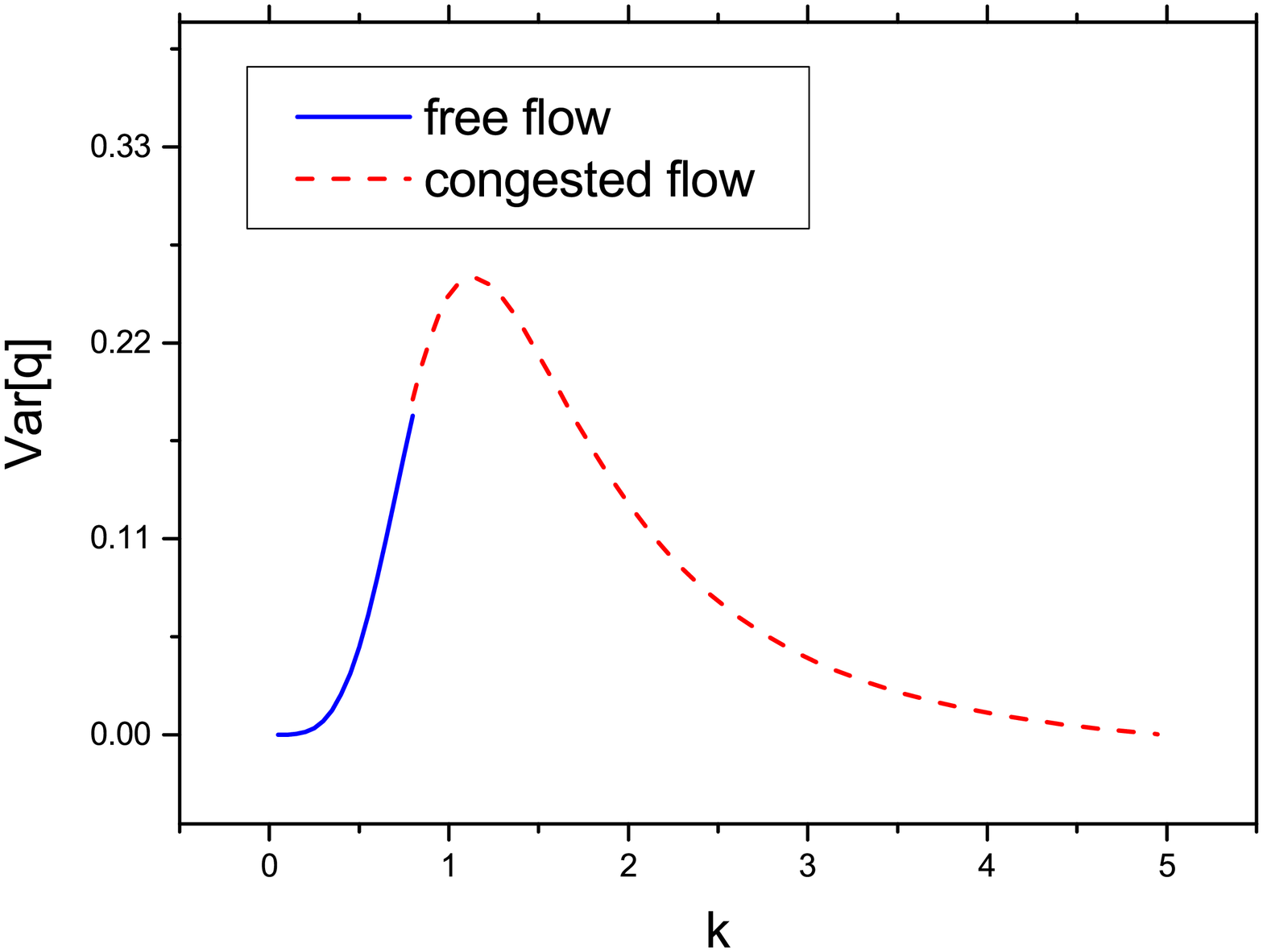} }
\end{minipage}
\\
\end{tabular}
\caption{Schematic fundamental diagram from the two-speed-state model including the maximal congestion density $k_{max}$.
We have use the same parameterization in Fig.\ref{fc-model} for the free flow phase and adopted $k_{max}=5$.
Left: the flow as a function of concentration, where capacity drop occurs at $k_c^{(1)}$. Right: variance of the flow as a function of concentration, where the connection between the two phases is quite smooth.}
\label{fc-model2}
\end{figure} 

In Fig.\ref{fc-model}, we show the fundamental diagram of the modified model. 
At the point of maximal flow where $k=k_c^{(1)}$, it is not difficult to see that there is a difference between the flow in the free flow phase and that in congested phase, and the latter is smaller by an amount given by

\begin{eqnarray}
\Delta q &\equiv& {\textrm {E}}[q^{(FT)}(k_c^{(1)})]- {\textrm {E}}[q^{(HCT)}(k_c^{(1)})]\nonumber \\
&=&  \frac{p_{11}v_2 k}{p_{11}+p_{22}L^{\alpha-1} \left( \frac{1}{\alpha-1}\frac{p_{11}}{p_{22}} \right)}-\frac{p_{11}v_2 k}{p_{11}+p_{22}L^{\alpha-1} \left( \frac{1}{\alpha-1}\frac{p_{11}}{p_{22}} \right)\left(\frac{1}{1-\frac{k}{k_{max}}}\right)}
\end{eqnarray}  
which corresponds to the capacity drop from the free flow to the congested flow. It can also be inferred from the plot that the variance evolves almost continuously and reaches its peak in the congested flow phase as observed experimentally, 
and the overall curve is not much affected by the inclusion of $k_{max}$. The main feature of the variance remains mostly unchanged. Since we are only dealing with the stable solution of the EoM, the congested flow in this study is only related to the stable homogeneous congested flow.

\section{Model calibration and data analysis}

In this section, the parameters of our model are adjusted to reproduce the observed flow-concentration data and its variance.
We make use of the data from I-80 freeway \citep{traffic-flow-data-08}, 
collected by the Federal Highway Administration (FHWA) and the Next Generation Simulation (NGSIM) program in the San Francisco Bay area in Emeryville, CA, on April 13, 2005.
The data was collected by seven synchronized digital video cameras and the vehicle trajectories were recorded. 
The study area was approximately 500 meters (1,640 feet) in length and consisted of six freeway lanes, including a high-occupancy vehicle (HOV) lane. 
In our calculations, we only take into consideration the automobiles (vehicle type 2 in the data set).

\begin{center}
\begin{figure}[!htb]
\centerline{\includegraphics*[width=11cm]{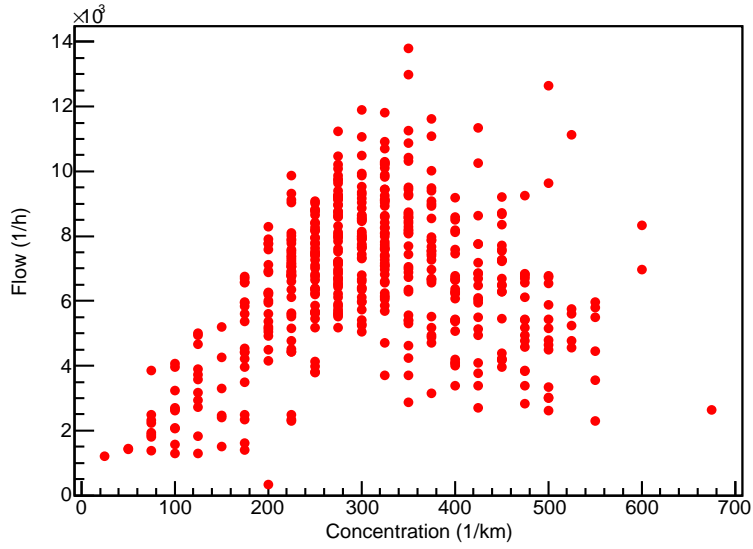} }
\caption{Observed fundamental diagram from the I-80 freeway in the San Francisco Bay area in Emeryville.
Here each data point corresponds to the measured flow and vehicle density obtained at different instants but at fixed locations. 
The data set used in this plot contains a total of 1,000 points.}
\label{fc-data}
\end{figure} 
\end{center}

\begin{table}[!hbt] 
\caption{Parameters of the two speed version of the model} 
\medskip
\begin{center}
\begin{tabular}{c|c|c|c|c|c|c}
\hline\hline
Parameters & $p_{11}$ & $p_{22}$ & $v_1$ & $v_2$ & $L$ & $\alpha$\\ 
\hline
P1&12.53 & 0.03 & 0.000012 & 66.74 & 0.105 & 1.898 \\
\hline
\end{tabular}
\end{center}
\label{tb1}
\end{table}

\begin{table}[!hbt] 
\caption{Parameters of the three speed version of the model} 
\medskip
\begin{center}
\begin{tabular}{c|c|c|c|c|c|c|c|c|c}
\hline\hline
Parameters & $p_{12}$ & $p_{13}$ & $p_{21}$ & $p_{23}$ & $p_{31}$ & $p_{32}$ & $v_1$ & $v_2$ & $v_3$ \\ 
\hline
P2& 2.11 & 0.000206 & 0.643 & 1.723 & 1.869 & 0.760 & 1.019 & 19.31 & 65.15 \\
\hline
\end{tabular}
\end{center}
\begin{center}
\begin{tabular}{c|c|c|c}
\hline\hline
$L$ & $\alpha_{12}$ & $\alpha_{13}$& $\alpha_{23}$\\ 
\hline
0.792 & 2.88 & 0.03 & 2.75 \\
\hline
\end{tabular}
\end{center}
\label{tb2}
\end{table}

In Fig.\ref{fc-data}, we show the observed fundamental diagram from the I-80 freeway in the San Francisco Bay area in Emeryville, CA, on April 13, 2005.
Each data point in the plot corresponds to the measured flow and vehicle density obtained for a fixed spatial interval at a given instant. 
The vehicle density is calculated by dividing the total number of vehicles by the size of the spatial interval which is taken as 40 meters in the calculation. 
The size of the above interval should be small enough to be sensitive to any traffic congestion and big enough to have reasonable resolution on the density axis of the plot.
In practice, several values were tested until the resulting plot satisfies both criterions.
The flow is obtained by the product of the average speed of all the vehicles within the interval and the vehicle density.

\begin{figure}[!htb]
\begin{tabular}{cc}
\begin{minipage}{200pt}
\centerline{\includegraphics*[width=8cm]{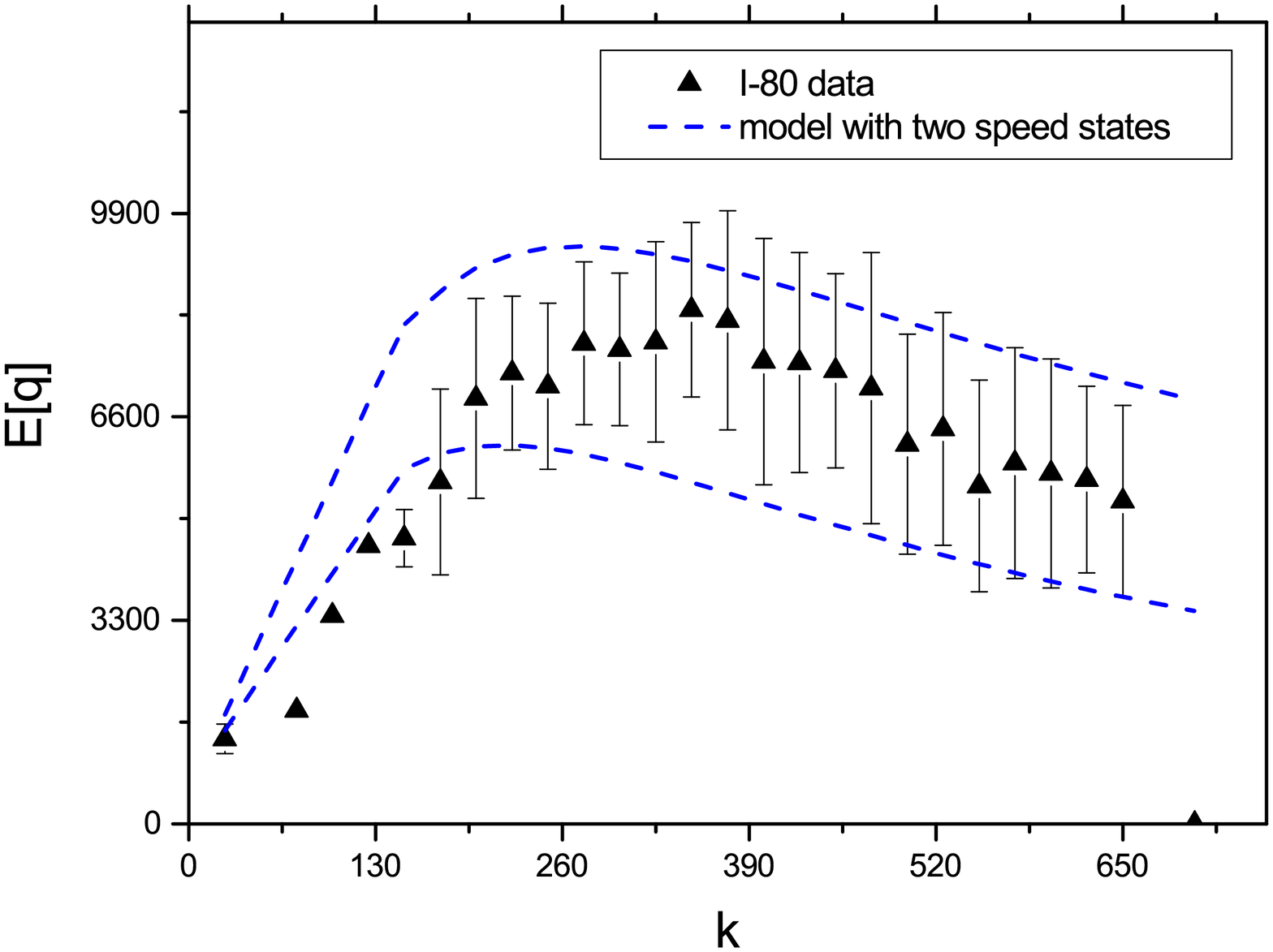} }
\end{minipage}
&
\begin{minipage}{200pt}
\centerline{\includegraphics*[width=8cm]{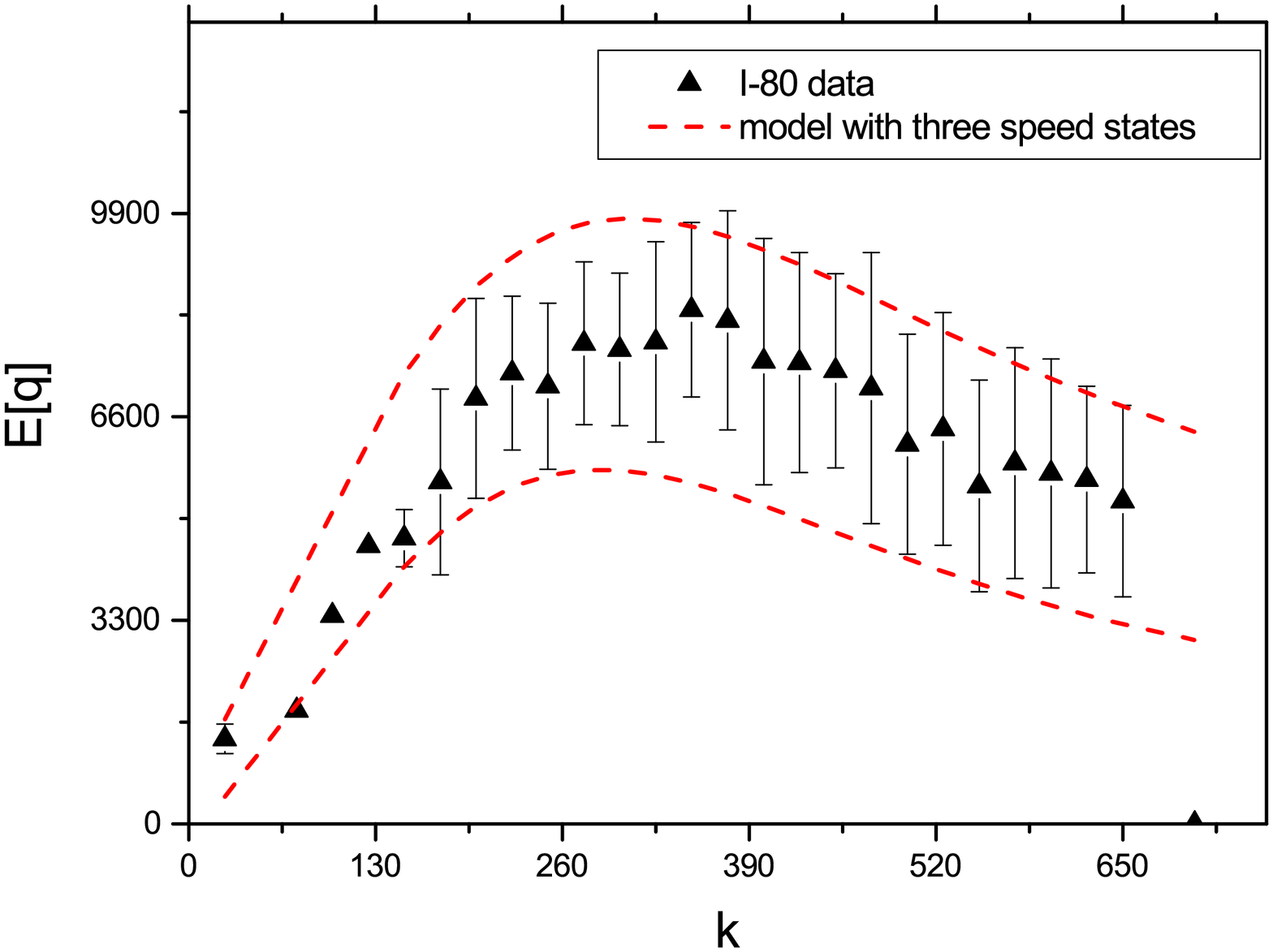} }
\end{minipage}
\\
\begin{minipage}{200pt}
\centerline{\includegraphics*[width=8cm]{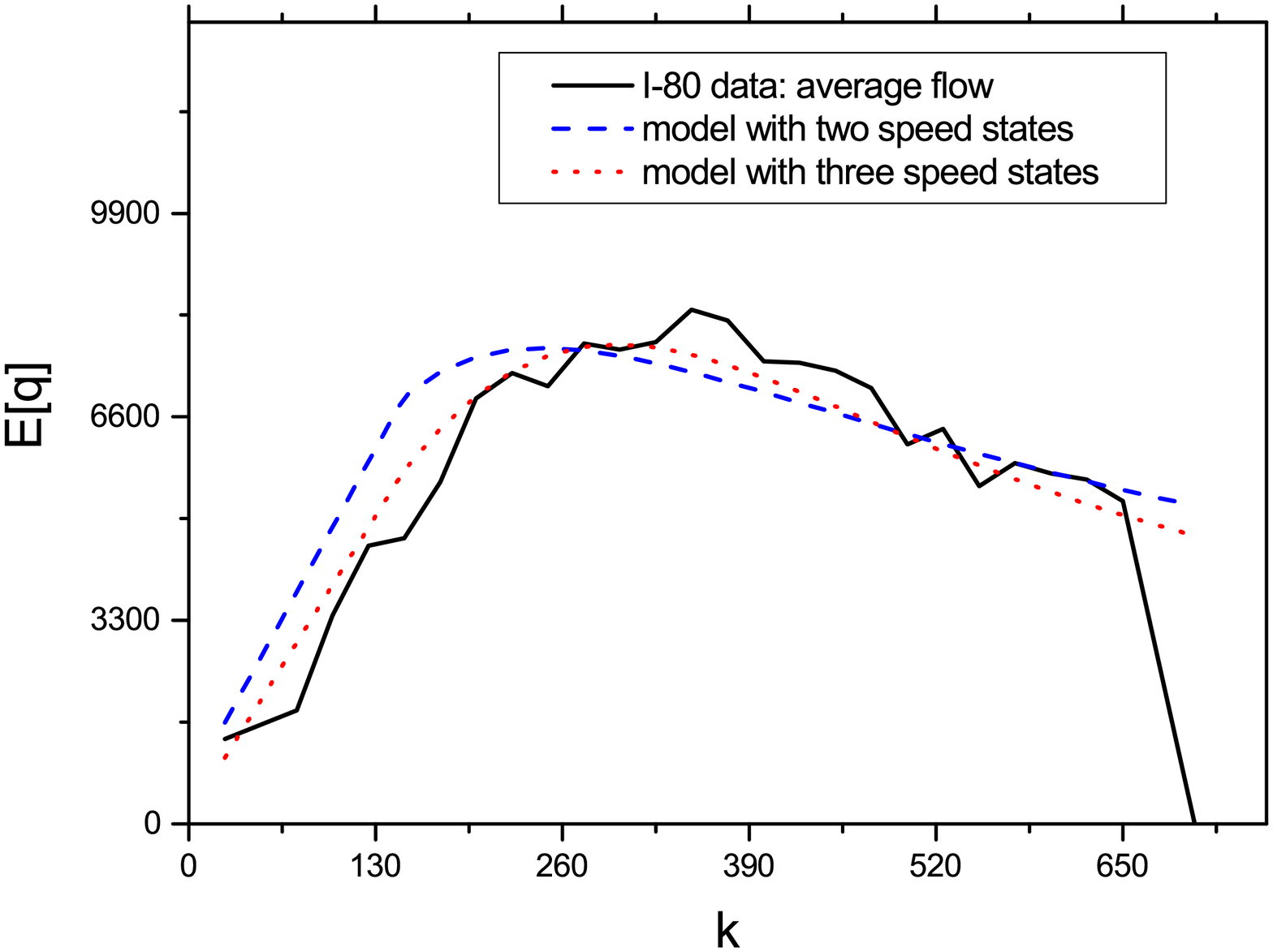} }
\end{minipage}
&
\begin{minipage}{200pt}
\centerline{\includegraphics*[width=8cm]{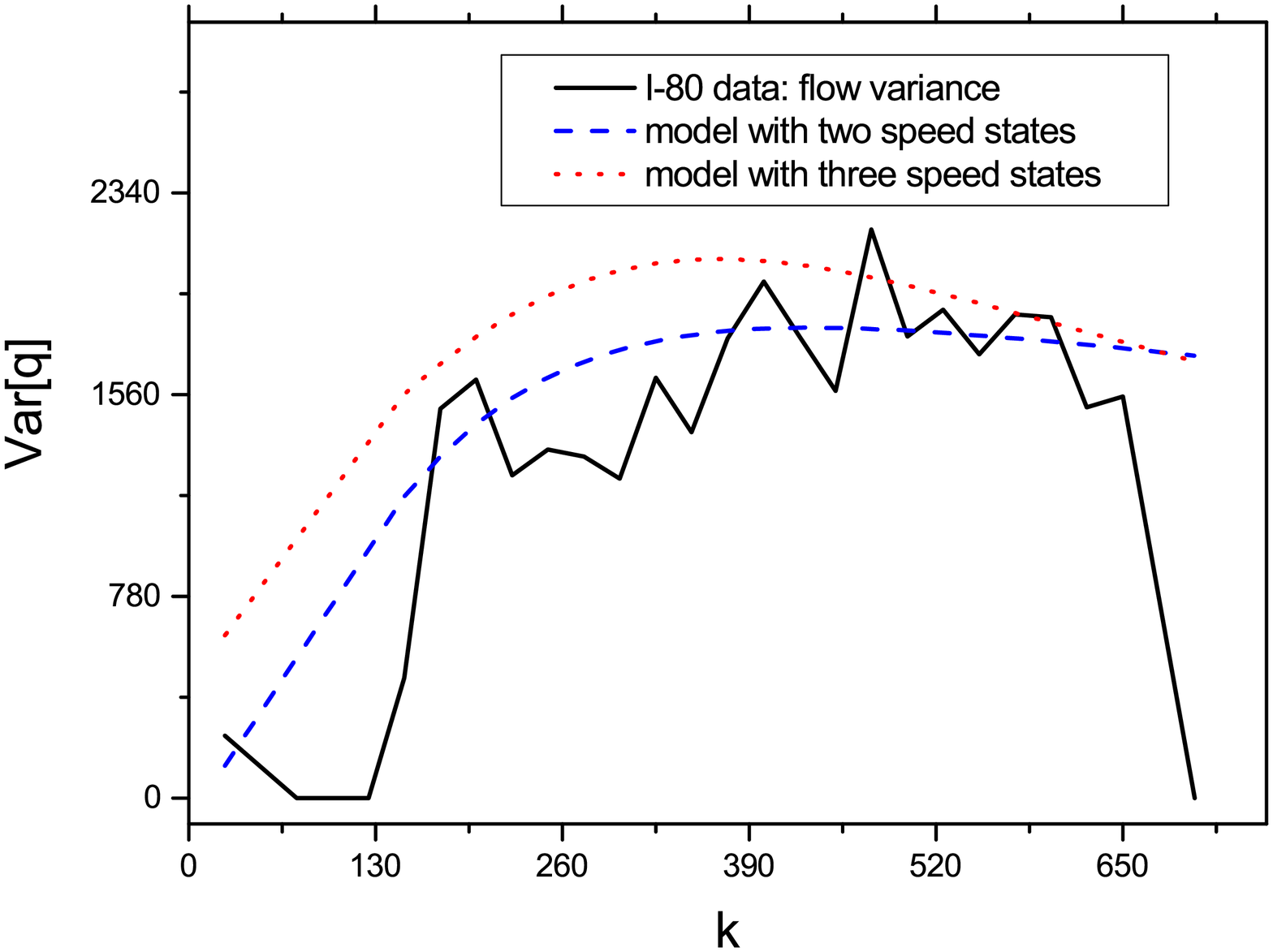} }
\end{minipage}
\\
\end{tabular}
\caption{The resulting fundamental diagram from two simplified versions of the model compared to the data in Fig.\ref{fc-data}.
Top left: the flow and flow variance curve by using the two speed version of the model. Top right: the same as the top left plot but using the three speed version of the model.
Bottom left: the expected value of the flow in comparison with two versions of the model. Bottom right: the variance of the flow in comparison with two versions of the model.}
\label{fc-fit}
\end{figure}

The above data is compared in Fig.\ref{fc-fit} with the results of two simplified versions of our model, where one considers two speed states and three speed states respectively.
The motivation of presenting the results of the three speed version of the model is to show explicitly that it is able to give a better description of the data.
We relegate the detailed formulae of the model with three speed states to the Appendix II,
and present the plots obtained by using the parameters in Tables \ref{tb1} and \ref{tb2}, where chi-square fitting was employed.
One sees that the model reproduces the main features observed in the fundamental diagram: 
flow increases from zero when the density of the vehicles increases, it hits the maximum then starts to decrease;
meanwhile, the flow variance also increases from zero with increasing density while the traffic starts to build up, it attains its maximum at a bigger density value than that of the flow.
The above calibrations did not consider the maximal congestion density $k_{max}$. Since the capacity drop in the data is not significant, the inclusion of $k_{max}$ does not yield much difference (not shown here).
It is worth noting that the model parameters obtained from calibration may vary when applying to a different traffic system.
However, the main features of the fundamental diagram discussed in the previous section are independent of any particular choice of parameters and therefore remain valid.

In the comparison between the two different versions of the model, the version considering three speed states reproduces the data better, even though the main characteristics of the fundamental diagram were reproduced reasonably well by both versions.
It is quite intuitive to understand that the version with more speed states is closer to the reality and therefore provides a better description of the observations.  
Taking into account the fact that the model only considers a few speed states, one may argue that it captures essence of the fundamental diagram as well as its variance.
We note in the above numerical calculations, the parameter $k_{max}$ was not considered.
It is intriguing to do further analysis by using more data sets from different highway systems.

\section{Conclusions remarks}

In this work, an alternative stochastic transport model is proposed to calculate the fundamental diagram of traffic flow and its variance. 
In order to show the physical content of our approach more transparently, 
we focus on a simplified version of our model with minimal number of parameters 
by following the spirit of other authors \citep{traffic-flow-review-07,traffic-flow-micro-07,traffic-flow-cellular-05,traffic-flow-micro-16,traffic-flow-micro-17}.
It is shown that even in this two-speed-state model, the stochastic nature of the model helps to capture the main features of the observed wide scattering of the flow-concentration data.
To describe the transition from ``free flow" to ``congested flow", 
the current model does not introduce different rules for different regions of the flow. 
The ``congested phase" in our model comes out naturaly as the
variance of the traffic flow grows with increasing flow, by solving a uniform set of SDE.
The model is then put to the test by calibration for the observed traffic flow data on the I-80 freeway from the NGSIM program, 
where both the fundamental diagram and its variance are reasonably reproduced.

SDE finds many applications in applied mathematics, and traffic congestion is such a phenomenon that contains within itself the nature of ``randomness". 
It follows that the study of the stochastic nature of traffic flow may provide valuable insight in our understanding of congestion flow.
On the practical side it may have significant impact on the ``uncertainty", or confidence interval, of the conclusions drawn by other deterministic traffic models, which may have important practical implications in policy making.
In fact, this is the exact motivation of our work. 
As mentioned above, the nature of ``randomness" or ``stochastic process" in the traffic flow has been studied by many authors.
Macroscopic \citep{traffic-flow-phenomenology-02,traffic-flow-phenomenology-03,traffic-flow-phenomenology-04,traffic-flow-phenomenology-05} as well as phenomenological models \citep{traffic-flow-phenomenology-06,traffic-flow-phenomenology-07,traffic-flow-phenomenology-08} usually do not depend on the details of the equation of motion of individual vehicles on the microscopic scale, where ``randomness" can be introduced into the collective behavior of the vehicles on a macroscopic level and one obtains the resulting hydrodynamic equation. 
In the present approach the uncertainty is planted on the microscopic level. 
It is different from the microscopic stochastic approaches \citep{traffic-flow-micro-16,traffic-flow-cellular-06} which involves discrete time interval, the SDE deals with continuous time variable and its derivatives.
However, the SDE modeling approach (eg. \citep{traffic-flow-micro-15}) has not been sufficiently explored, where the mathematical concept of ``signal noise" (such as white noise or Brown noise which are more realistic in modeling random process by their nature) is employed and most studies are usually heavily based on numerical solutions. 
In the present work, the approach is based on an analytic solution for the expected value and variance of SDE in Ito interpretation.
In terms of the physical system under investigation, the car-following \citep{traffic-flow-micro-13,traffic-flow-micro-14,traffic-flow-micro-18} and the cellular automaton models \citep{traffic-flow-cellular-01,traffic-flow-cellular-02,traffic-flow-cellular-03,traffic-flow-cellular-04,traffic-flow-cellular-05} usually incorporate specific rules for each of the different traffic scenarios which are based on empirical observations.
Our approach, on the other hand, is based on a special version of the gas-kinetic theory.
It is understood that, in the spirit of Liouville's theorem, most of the times the equation of motion of the system can be cast into the form of a transport equation where the system dynamics is mapped into the coefficients of transition rate.
In this context, the mathematical form of the current model is independent of any specific traffic scenario, which may be used to describe traffic evolution under many different traffic conditions.

Though the stochastic transition term plays an important role in this study, we have deliberately avoided the discussion of some mathematical aspects of our approach.
For instance, the simple form of white noise is adopted to describe the uncertainty in vehicle transitions, where the corresponding SDE is interpreted in terms of It{\^o} formulae.
However, it is well known that there are various forms of stochastic noises in stochastic calculus, and in fact it is not clear if other forms of stochastic noise might be more appropriate for the description of the physical system.
We have not discussed whether Stratonovich stochastic calculus could be more convenient to our investigation either.
However, we argue that these features are not the focus of the present study, because the main goal of this work is to study qualitatively the effect of the stochastic transition by employing a simple model with the minimal number of parameters.
Therefore, as a first step, it is of higher priority to reproduce the main feature of the observed flow-concentration curve as well as its variance. 
In order to understand the physical content of the model more transparently, it is worthwhile to simplify the mathematics.
One notes that in our approach only the homogeneous solution is studied and the vehicle density $k$ may approach infinity. 
In the real world, however, owing to the physics size of the vehicle, the density always has an upper limit.
Therefore when comparing to the data, the discussion should be restrained in the physical region where realistic values of $k$ apply.

One important aspect which has not been thoroughly discussed in this work is the stability of the traffic system.
As mentioned above, it was understood by many authors that the traffic congestion is closely connected to the instability of small perturbations\citep{traffic-flow-hydrodynamics-04,traffic-flow-micro-07,traffic-flow-micro-08,traffic-flow-micro-09}.
It involves two concepts. First, as discussed at the end of Section II, the problem of stability of deterministic differential equation is closely connected with that of traffic flow.
Second, since our approach itself involves a mathematical description of uncertainty, it is quite natural to ask whether the problem of stability of SDE \citep{stochastic-difeq-stability-01,stochastic-difeq-stability-02,stochastic-difeq-stability-03,stochastic-difeq-stability-04} may further complicate the matter.
It is of course an interesting topic worthy of further exploration.


\section{Acknowledgments}

Wei-Liang Qian is thankful for valuable discussions with Marcio Maia Vilela, Jos\'e Aquiles Baesso Grimoni, Pasi Huovenin, Kai Lin and Yogiro Hama.
We are grateful to the Federal Highway Administration (FHWA) and the Next Generation Simulation (NGSIM) program who provide the data, which are licensed under the terms specified by the modified creative commons license attribute 2.0 (http://creativecommons.org/licenses/by/2.0/).
This paper is greatly benefitted from Dr. Ted William Grant who carefully read the manuscript and gave us a lot of helpful advice on written English.
We acknowledge the financial support from Funda\c{c}\~ao de Amparo \`a Pesquisa do Estado de S\~ao Paulo (FAPESP), Funda\c{c}\~ao de Amparo \`a Pesquisa do Estado de Minas Gerais (FAPEMIG), Conselho Nacional de Desenvolvimento Cient\'{\i}fico e Tecnol\'ogico (CNPq), and Coordena\c{c}\~ao de Aperfei\c{c}oamento de Pessoal de N\'ivel Superior (CAPES).

\section{Appendix I: A derivation of the variance of two-speed-state model}

Here we derive the results used in section III together with some additional remarks. To start, it is convenient to write down the equation of motion of our two-speed-state model as 
\begin{eqnarray}
{dn_1} &=& -p_{11}n_1{dt} + p_{12}n_2 N^\alpha {dt} - \sqrt{p_{11}n_1} dB_1+\sqrt{p_{12}n_2 N^\alpha} dB_2 \nonumber \\
{dn_2} &=& -p_{22}n_2 N^\alpha {dt}+ p_{21}n_1  {dt} - \sqrt{p_{22}n_2 N^\alpha}dB_2 + \sqrt{p_{21}n_1} dB_1
\label{bte2b}
\end{eqnarray}
where $B_1$ and $B_2$ are independent Brownian motions. To evaluate the variance, one needs to calculate the expected value of $n_i^2$. By making use of ${d(n_1^2)} = 2n_1dn_1+(dn_1)^2$ and ${d(n_1n_2)}=n_1dn_2+n_2dn_1$, it is straightforward to show that
\begin{eqnarray}
{d(n_1^2)} &=& \left[(-2p_{11}-2p_{12}N^\alpha)n_1^2+(2p_{12}N^{\alpha+1}-p_{12}N^\alpha+p_{11})n_1+p_{12}N^{\alpha+1}\right]dt \nonumber \\
&&- 2\sqrt{p_{11}n_1^3} dB_1+2n_1\sqrt{p_{12}N^\alpha(N-n_1)} dB_2 \nonumber \\
{d(n_2^2)} &=& \left[(-2p_{22}N^\alpha-2p_{21})n_2^2+(p_{22}N^\alpha+2p_{21}N-p_{21})n_2+p_{21}N\right]{dt} \nonumber \\
&&- 2\sqrt{p_{22}n_2^3 N^\alpha}dB_2 + 2n_2\sqrt{p_{21}(N-n_2)} dB_1 \nonumber \\
{d(n_1n_2)} &=& \left[p_{21}n_1^2-(p_{11}+p_{22}N^\alpha)n_1n_2+p_{12}N^\alpha n_2^2\right]dt \nonumber \\
&& (n_1\sqrt{p_{21}n_1} - n_2\sqrt{p_{11}n_1} ) dB_1+ (n_2\sqrt{p_{12}n_2 N^\alpha} - n_1\sqrt{p_{22}n_2 N^\alpha}) dB_2
\end{eqnarray}
For steady state, one has
\begin{eqnarray}
{\textrm E}[n_1^{2(\infty)}] &=& \frac{p_{22}N^{\alpha+1}(2p_{22}N^{\alpha+1}-p_{22}N^\alpha+p_{11})}{2(p_{11}+p_{22}N^\alpha)^2} +\frac{p_{22}N^{\alpha+1}}{2(p_{11}+p_{22}N^\alpha)} \nonumber \\
{\textrm E}[n_2^{2(\infty)}] &=&\frac{p_{11}N(p_{22}N^{\alpha}+2p_{11}N-p_{11})}{2(p_{11}+p_{22}N^\alpha)^2} +\frac{p_{11}N}{2(p_{11}+p_{22}N^\alpha)} \nonumber \\
{\textrm E}[{n_1n_2}^{(\infty)}] &= & N{\textrm E}[n_1^{(\infty)}]-{\textrm E}[n_2^{2(\infty)}]=\frac{p_{11}p_{22}N^{\alpha}(N-1)}{(p_{11}+p_{22}N^\alpha)^2} 
\end{eqnarray}
where one recalls the results obtained before in Eq.(\ref{expectedsteady})
\begin{eqnarray}
{\textrm E}[n_1^{(\infty)}] &=& \frac{p_{22}N^{\alpha+1}}{p_{11}+p_{22}N^\alpha} \nonumber \\
{\textrm E}[n_2^{(\infty)}] &=& \frac{p_{11}N}{p_{11}+p_{22}N^\alpha}  \nonumber
\end{eqnarray}
and makes the substitutions $p_{11}=p_{21}$ and $p_{22} = p_{12}$, and one is readily to verify the following identities
\begin{eqnarray}
{\textrm E}[n_1^{2(\infty)}]+{\textrm E}[n_2^{2(\infty)}] &=& N({\textrm E}[n_1^{(\infty)}]+{\textrm E}[n_2^{(\infty)}])-2{\textrm E}[{n_1n_2}^{(\infty)}] \nonumber \\
{\textrm E}[n_1^{2(\infty)}]-{\textrm E}[n_2^{2(\infty)}] &=& N({\textrm E}[n_1^{(\infty)}]-{\textrm E}[n_2^{(\infty)}]) \nonumber 
\end{eqnarray} 

Now, one is in the position to calculate the variances, which turn out to be quite simple in their forms
\begin{eqnarray}
{\textrm {Var}}[n_1^{(\infty)}]= {\textrm {Var}}[n_2^{(\infty)}] = - {\textrm {Cov}}[{n_1n_2}^{(\infty)}] = \frac{p_{11}p_{22}N^{\alpha+1}}{(p_{11}+p_{22}N^\alpha)^2} 
\end{eqnarray}
The above resulting expression is partly due to the fact that
\begin{eqnarray}
{\textrm {Var}}[n_1^{(\infty)}]+{\textrm {Var}}[n_2^{(\infty)}] +2 {\textrm {Cov}}[{n_1n_2}^{(\infty)}]={\textrm {Var}}[N^2] = 0 \nonumber \\
{\textrm {Var}}[n_1^{(\infty)}]={\textrm {Var}}[-n_1^{(\infty)}]={\textrm {Var}}[N-n_1^{(\infty)}]={\textrm {Var}}[n_2^{(\infty)}]
\end{eqnarray}

Putting all the pieces together, one obtains the measured variance of vehicle speed
\begin{eqnarray}
{\textrm {Var}}[v] &\equiv& {\textrm {Var}}\left[\frac{\sum_i [n_i^{(\infty)}] v_i}{\sum n_i}\right] \nonumber \\
&=& \left(\frac{v_1}{N}\right)^2 {\textrm {Var}}[n_1^{(\infty)}]+\left(\frac{v_2}{N}\right)^2 {\textrm {Var}}[n_2^{(\infty)}]+2\left(\frac{v_1v_2}{N^2}\right) {\textrm {Cov}}[n_1n_2] \nonumber \\
&=& \frac{(v_1-v_2)^2}{N^2}\frac{p_{11}p_{22}N^{\alpha+1}}{(p_{11}+p_{22}N^\alpha)^2}
\end{eqnarray}
By substituting the concentration $k$, one arrives at the expression of variance of flow
\begin{eqnarray}
{\textrm {Var}}[q] = \frac{(v_1-v_2)^2}{L^2}\frac{p_{11}p_{22}L^{\alpha+1}k^{\alpha+1}}{(p_{11}+p_{22}L^\alpha k^\alpha)^2} \nonumber
\end{eqnarray}
used in Section III.

\section{Appendix II: Formulae for the simplified model with three speed states}

Here we present the equation of motion of the three-speed-state model as well as its analytic solutions. The equation of motion reads
\begin{eqnarray}
{dn_1} &=& -p_{21}n_1{dt}-p_{31}n_1{dt} + p_{12}n_2 N^{\alpha_{12}} {dt}+ p_{13}n_3 N^{\alpha_{13}} {dt} \nonumber \\
 &-& \sqrt{p_{21}n_1} dB_{21}- \sqrt{p_{31}n_1} dB_{31}+\sqrt{p_{12}n_2 N^{\alpha_{12}}} dB_{12}+\sqrt{p_{13}n_3 N^{\alpha_{13}}} dB_{13} \nonumber \\
{dn_2} &=& p_{21}n_1{dt} -p_{32}n_2{dt}-p_{12}n_2N^{\alpha_{12}}{dt}+ p_{23}n_3 N^{\alpha_{23}}{dt}  \nonumber \\
 &-& \sqrt{p_{12}n_2 N^\alpha_{12}}dB_{12}- \sqrt{p_{32}n_2 }dB_{32} +\sqrt{p_{23}n_3 N^\alpha_{12}}dB_{23}+ \sqrt{p_{21}n_1} dB_{21}\nonumber \\
{dn_3} &=& p_{31}n_1{dt}+ p_{32}n_2  {dt}- p_{13}n_3N^{\alpha_{13}}  {dt}-p_{23}n_3N^{\alpha_{23}}  {dt} \nonumber \\
&-& \sqrt{p_{23}n_3 N^\alpha_{23}}dB_{23}- \sqrt{p_{13}n_3 N^\alpha_{13}}dB_{13} + \sqrt{p_{31}n_1} dB_{31}+ \sqrt{p_{32}n_2} dB_{32}
\label{bte3}
\end{eqnarray}
It is noted that three $\alpha$ factors are introduced only to the transitions from high speed state to low speed state. The expected values are obtained after some calculations
\begin{eqnarray}
{\textrm E}[n_1] &=& \frac{bN}{a+b+c} \nonumber \\
{\textrm E}[n_2] &=& \frac{cN}{a+b+c} \nonumber \\
{\textrm E}[n_3] &=& \frac{aN}{a+b+c}
\label{avg3}
\end{eqnarray}
where
\begin{eqnarray}
a &=& p_{21}p_{32}+p_{31}p_{32}+p_{31}p_{12}N^{\alpha_{12}} \nonumber \\
b &=& p_{32}p_{13}N^{\alpha_{13}}+p_{12}N^{\alpha_{12}}p_{13}N^{\alpha_{13}}+p_{12}N^{\alpha_{12}}p_{23}N^{\alpha_{23}} \nonumber \\
c &=& p_{21}p_{13}N^{\alpha_{13}}+p_{21}p_{23}N^{\alpha_{23}}+p_{31}p_{23}N^{\alpha_{23}}
\label{abc}
\end{eqnarray}
After some lengthy but still manageable calculations, one obtains the following system of equations for the variances
\begin{eqnarray}
(2C+2A){\textrm E}[n_1^2] &=& (2CN+A-C){\textrm E}[n_1]+(B-C){\textrm E}[n_2]+(2B-2C){\textrm E}[n_1n_2]+NC \nonumber \\
(2F+2G){\textrm E}[n_2^2] &=& (D-G){\textrm E}[n_1]+(2NG+F-G){\textrm E}[n_2]+(2D-2G){\textrm E}[n_1n_2]+NG \nonumber \\
(F+A+G+C){\textrm E}[n_1n_2] &=& (D-G){\textrm E}[n_1^2]+(B-C){\textrm E}[n_2^2]+(GN-D){\textrm E}[n_1]\nonumber \\
&+&(CN-I){\textrm E}[n_2]
\label{var31}
\end{eqnarray}
where
\begin{eqnarray}
A &=& p_{21}+p_{31} \nonumber \\
B &=& p_{12}N^{\alpha_{12}} \nonumber \\
C &=& p_{13}N^{\alpha_{13}} \nonumber \\
D &=& p_{21} \nonumber \\
E &=& p_{31} \nonumber \\
F &=& p_{32}+p_{12}N^{\alpha_{12}} \nonumber \\
G &=& p_{23}N^{\alpha_{23}} \nonumber \\
H &=& p_{32} \nonumber \\
I &=& p_{12}N^{\alpha_{12}}
\label{ABCDEFGHI}
\end{eqnarray}
and
\begin{eqnarray}
{\textrm Var}[n_1] &=& {\textrm E}[n_1^2]-({\textrm E}[n_1])^2 \nonumber \\
{\textrm Var}[n_2] &=& {\textrm E}[n_2^2]-({\textrm E}[n_2])^2 \nonumber \\
{\textrm Var}[n_3] &=& {\textrm Var}[n_1+n_2]= {\textrm Var}[n_1]+{\textrm Var}[n_2]+2{\textrm Cov}[n_1n_2] \nonumber \\
{\textrm Cov}[n_1n_2]&=& {\textrm E}[n_1n_2]-{\textrm E}[n_1]{\textrm E}[n_2] \nonumber \\
{\textrm Cov}[n_1n_3]&=& -{\textrm Var}[n_1]-{\textrm Cov}[n_1n_2] \nonumber \\
{\textrm Cov}[n_2n_3]&=& -{\textrm Var}[n_2]-{\textrm Cov}[n_1n_2]
\label{var32}
\end{eqnarray}
Putting all the pieces together, one gets the expression for the flow and its variance
\begin{eqnarray}
{\textrm E}[q] &=& \frac{N}{L}\frac{n_1v_1+n_2v_2+n_3v_3}{N} \\
{\textrm Var}[q] &=& \frac{v_1^2}{L^2}{\textrm Var}[n_1]+\frac{v_2^2}{L^2}{\textrm Var}[n_2]+\frac{v_3^2}{L^2}{\textrm Var}[n_3]\nonumber \\
&+&\frac{v_1v_2}{L^2}{\textrm Cov}[n_1n_2]+\frac{v_1v_3}{L^2}{\textrm Cov}[n_1n_3]++\frac{v_2v_3}{L^2}{\textrm Cov}[n_2n_3]
\label{qvar3}
\end{eqnarray}

\bibliographystyle{ormsv080}

\bibliography{references_qian}{}

\end{document}